\documentclass[
    ,draft            % use draft while you are working on the paper
  ]
  {aipproc}

\layoutstyle{6x9}

\begin{document}

\title{SNe 2005ay and 2005cs: Two interesting II-Plateau events}
%shape and shift of, 32.70.Jz, 33.70.Jg

%<Replace this text with PACS numbers; choose from this list:
%                \texttt{http://www.aip..org/pacs/index.html}>}

\author{I.J. Danziger}{
  address={Osservatorio Astronomico di Trieste, Via G.B.Tiepolo 11 - I-34131
	 Trieste - Italy }
}
\author{A. Elmhamdi}{
 address={<common address for author1 and author2>}
}
\author{D.Yu. Tsvetkov }{
 address={<Sternberg State Astronomical Institute, Universitetskii pr. 13, 
              119992, Moscow, Russia>}
}
\author{A.A. Volnova }{
 address={<common address for author3 and author4>}
}
\author{A.P. Shulga }{
 address={<common address for author3 and author5>}
}

\begin{abstract}
 The $U,B,V,R,I$ light curves of the type IIP supernovae (SNe IIP)
 SN 2005ay and SN 2005cs, and one spectrum for SN 2005cs, are
 presented and analyzed. We found both events to be fainter than 
 the average SN IIP, with SN 2005cs showing slight  
 brightening in the second half of plateau stage in the
 $V,R,I$ bands and a low expansion velocity.
 The effects of two different plausible distance moduli on the
 derived physical parameters of SN 2005ay are considered.
 Two approaches are used
  to estimate the expansion velocities at the middle of
 the plateau phase. Based on empirical analytical models
 we derived constraints on the progenitor properties.
 The amounts of the  ejected $^{56}$Ni are also recovered.

\end{abstract}

\maketitle

%%%%%%%%%%%%%%%%%%%%%%%%%%%%%%%%%%%%%%%%%%%%
%% MAINMATTER
%%%%%%%%%%%%%%%%%%%%%%%%%%%%%%%%%%%%%%%%%%%%

\section{Data and Main parameters}
 {\bf SN 2005ay:}
 SN 2005ay was discovered, in the Sc galaxy NGC 3938,   
 on March 27 by D. Rich at magnitude 15.6 on CCD frames taken with 
 a 0.31-m reflector \cite{Rich05}.  
 A spectrum obtained on March 29.98 UT with the Calar Alto 2.2-m telescope 
 indicated a type-II SN nature, taken soon after the explosion \cite{Tau05}.

 Fig.1 (left) displays the $U,B,V,R,I$ light curves of SN 2005ay. Data
 of SN IIP 1999gi \cite{Leo02} are shown
 for comparison. 
 We shifted the light curves of SN 1999gi along the x-axis to
 reach the best agreement at the stage of the steep decline after the plateau.
 After a clear plateau phase,
 the light curves show a rapid drop of about 2 mag, followed
 by a linear tail with an $R-$band decline rate very close to the standard value
 of 0.01 mag/day, similar to the radioactive decay of $^{56}$Co.     
 The main estimated data are: 1- Explosion time: $\sim$March 23, 2005 
 (JD 2453453), 2- Total colour excess: $E(B-V)=0.1$, 3- Distance 
 Modulus: 30.82, based on the recession velocity of the host corrected
 for Local Group infall onto the Virgo Cluster as reported in the 
  ``LEDA'' database\footnote{http://leda.univ-lyon1.fr/}.
 However we note here that the host galaxy 
 NGC 3938 belongs to the $Ursa~ Major$ group. The spiral galaxy NGC 3982, which
 is also part of the group, has a Cepheid calibrated distance modulus
 of 31.71 \cite{Saha01}. If assuming the two 
 galaxies have this same distance, then SN 2005ay would
 be intrinsically more luminous, which would increase the synthesized 
 $^{56}$Ni mass by a factor of $\sim$2.26.\\ \\

 {\bf SN 2005cs:} 
The discovery of SN 2005cs, in the Sbc galaxy M51, was reported by 
 W. Kloehr on June 28 at a magnitude of about 14 \cite{Klo05}.
Modjaz et al. (2005)\cite{Mod05} reported the type II nature of the
 events based on the spectrum  obtained on June
30.23 UT with the F. L. Whipple Observatory 1.5-m telescope. 
Nothing was visible at this location on earlier frames 
 taken by W. Kloehr on May 11 and 26. 

 The photometry data of SN 2005cs, spanning more than a one year observations,
  are shown in Fig. 1(right), together with the typical SN IIP 1999em \cite{Elm03a}. 
 The type IIP nature of SN 2005cs is clear, with noticeable differences
 with SN 2005ay. 
 The early magnitudes from amateurs, taken by different and independent 
 obsevers, may suggest the presence
 of a narrow peak before the onset of the plateau in the $V$ 
 and $R$ bands at JD 2453551-552. The maximum light in $B$ band
 was reached on JD 2453553 with $B_{max}=14.5$ mag. 
 In the $V$ band, SN 2005cs
was brightest at a very early epoch, with $V_{max}=14.3$.
A similar behavior was observed for SN 1999em at the plateau stage,
but the increase in brightness was more pronounced for SN 2005cs.
The plateau stage lasted until about day JD 2453660,
and then the light curves display a steep decline with a drop of
 about 2.8 mag, then the exponential tail started.
 The main adopted data are: 1- Explosion time:  $\sim$June 27.3, 2005
 (JD 2453548.8), 2- Total colour excess:  $E(B-V)\simeq 0.14$ \cite{Mau05}, 
 3- Distance Modulus: $\mu=29.6$ \cite{Feld97,Mau05} 
%%%%%%%%%%%%%%%%%%%%%%%%%%%%%%%%%%
\begin{figure}
 \includegraphics[height=.3\textheight]{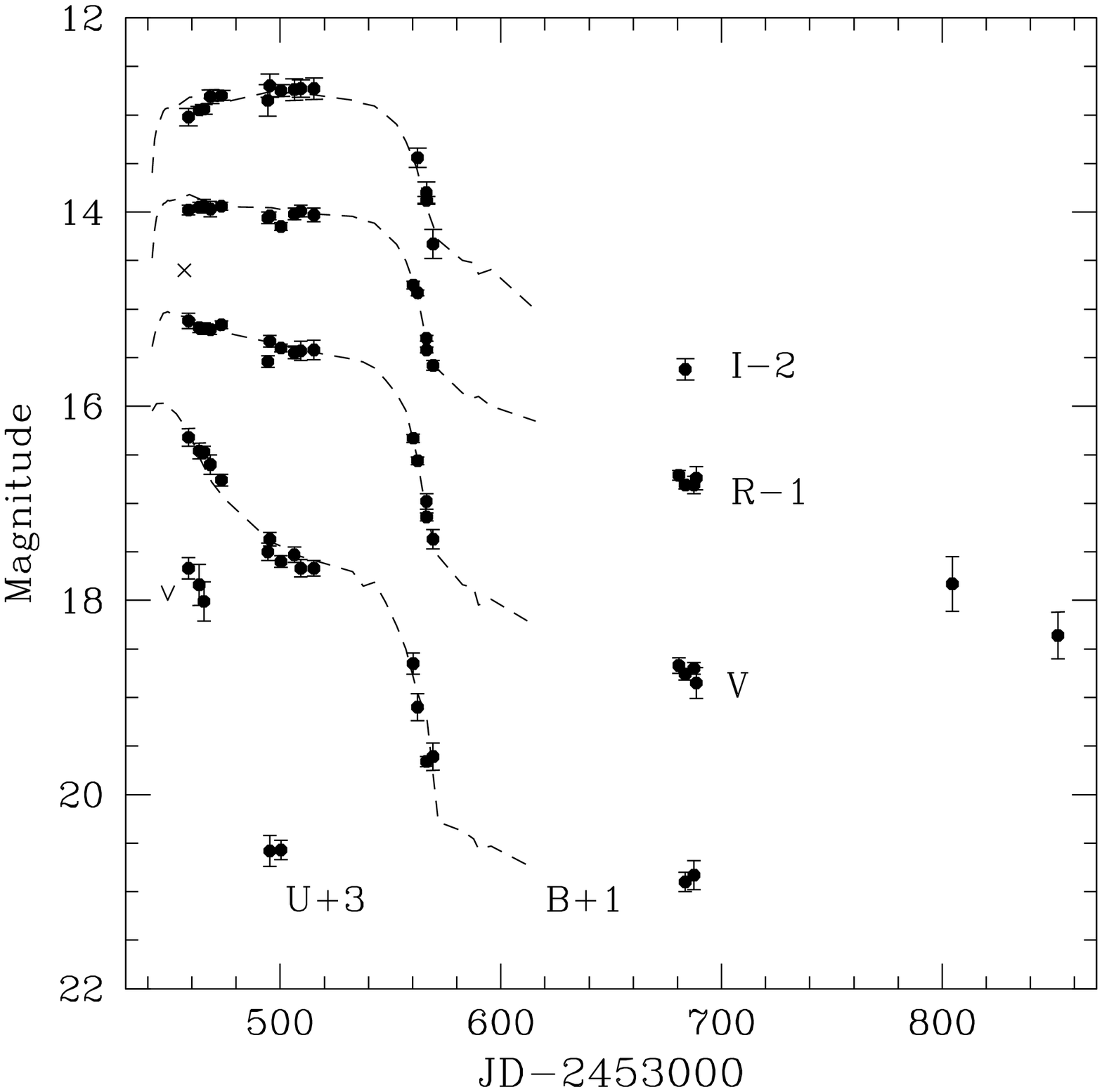}\includegraphics[height=.3\textheight]{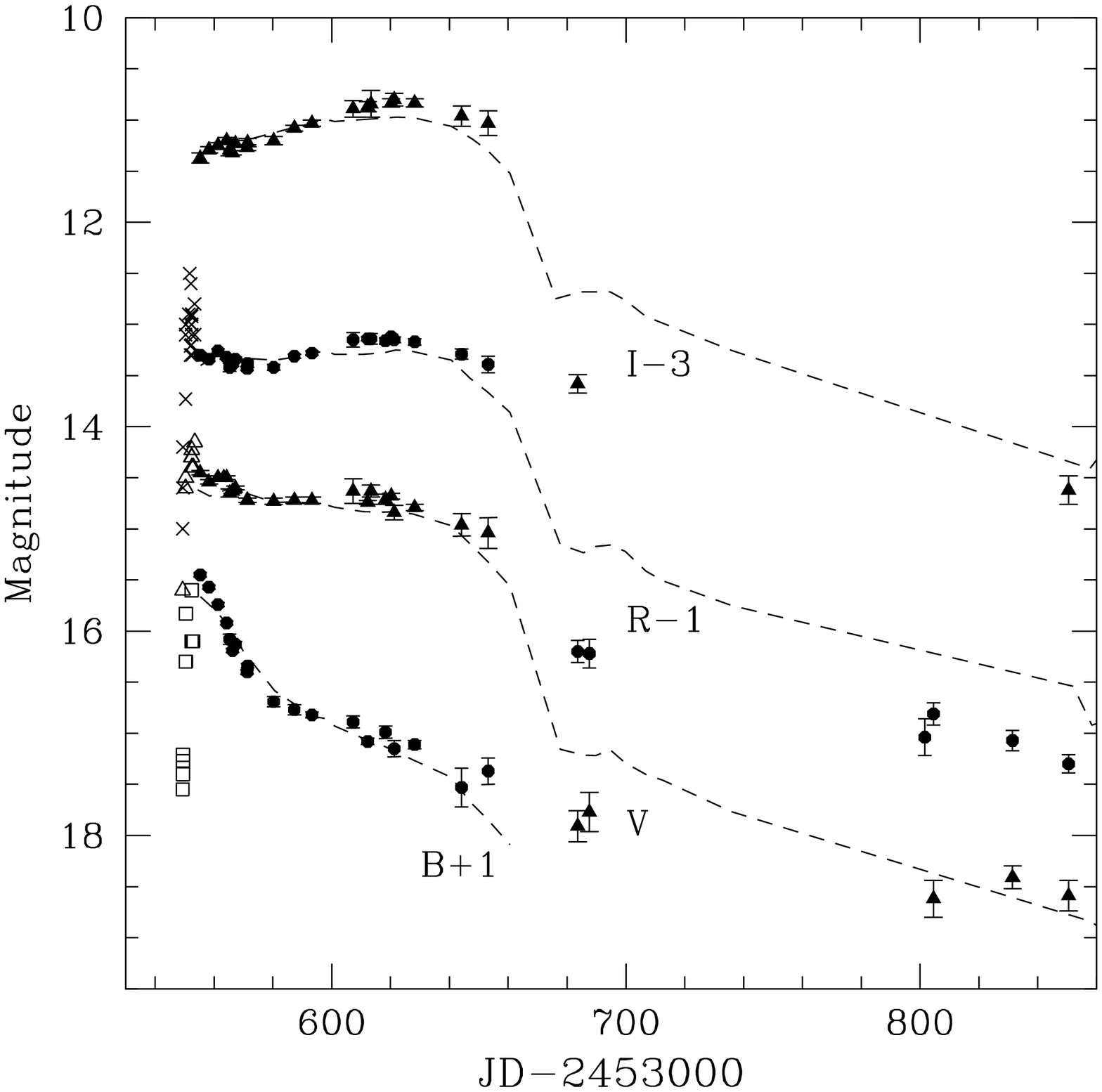}
  \caption{$Left$: $UBVRI$ light curves of SN 2005ay. Dots show
our data, cross is for the discovery magnitude by Rich (2005)\cite{Rich05},
the ``v'' mark is the upper limit from Yamaoka \& Itagaki (2005)\cite{Yam05}. 
The dashed lines are the light curves of SN 1999gi.
 $Right$: $BVRI$ light curves of SN 2005cs. Dots show
our data, crosses are for the observations of amateur astronomers.
The dashed lines are the light curves of SN 1999em.}

\end{figure}
%%%%%%%%%%%%%%%%%%%%%%%%%%%%%%%%%%

 The spectrum of SN 2005cs obtained on July 6 (JD 2453558.37)
 is presented in Fig. 2(left). A spectrum analysis was made by means
 of the synthetic code SYNOW \cite{Elm06}. 
 The velocities are found to be 
 significantly lower than the values for SN 1999em at similar epochs. The best 
 fit synthetic spectrum has a blackbody temperature
 of T$_{bb}=$ 14500 K, a photospheric velocity of 
 V$_{phot}=$ 5000 km s$^{-1}$,
 and including 7 ions, namely: H I, Fe II, Ca II, Mg II, Mg I, Sc II and Ni II.
 The H$\beta$ and H$\gamma$ features are
 prominent with typical P-Cygni profiles.

 The photometric measurements of the two SNe were made relative to 
 comparison stars using PSF-fitting technique,
 and in some cases using aperture photometry.
 The background of the host galaxies around the SNe 
 did not present any problems when the SNe were bright, but at late 
 stages it could introduce additional errors. We note here that image
 substruction technique is more appropriate at late phases, however
 since the SNe are still visible on our images, the subtraction technique
 is not possible. Thus, the magnitudes
 at late stages can be regarded as provisional, and they should be
 verified later when images without the SNe are obtained and
 used for image subtraction. 
%%%%%%%%%%%%%%%%%%%%%%%%%%%%%%%%%%
\section{Discussion \& Results}
%%%%%%%%%%%%%%%%%%%%%%%%%%%%%%%%%%
\begin{figure}
 \includegraphics[height=.3\textheight]{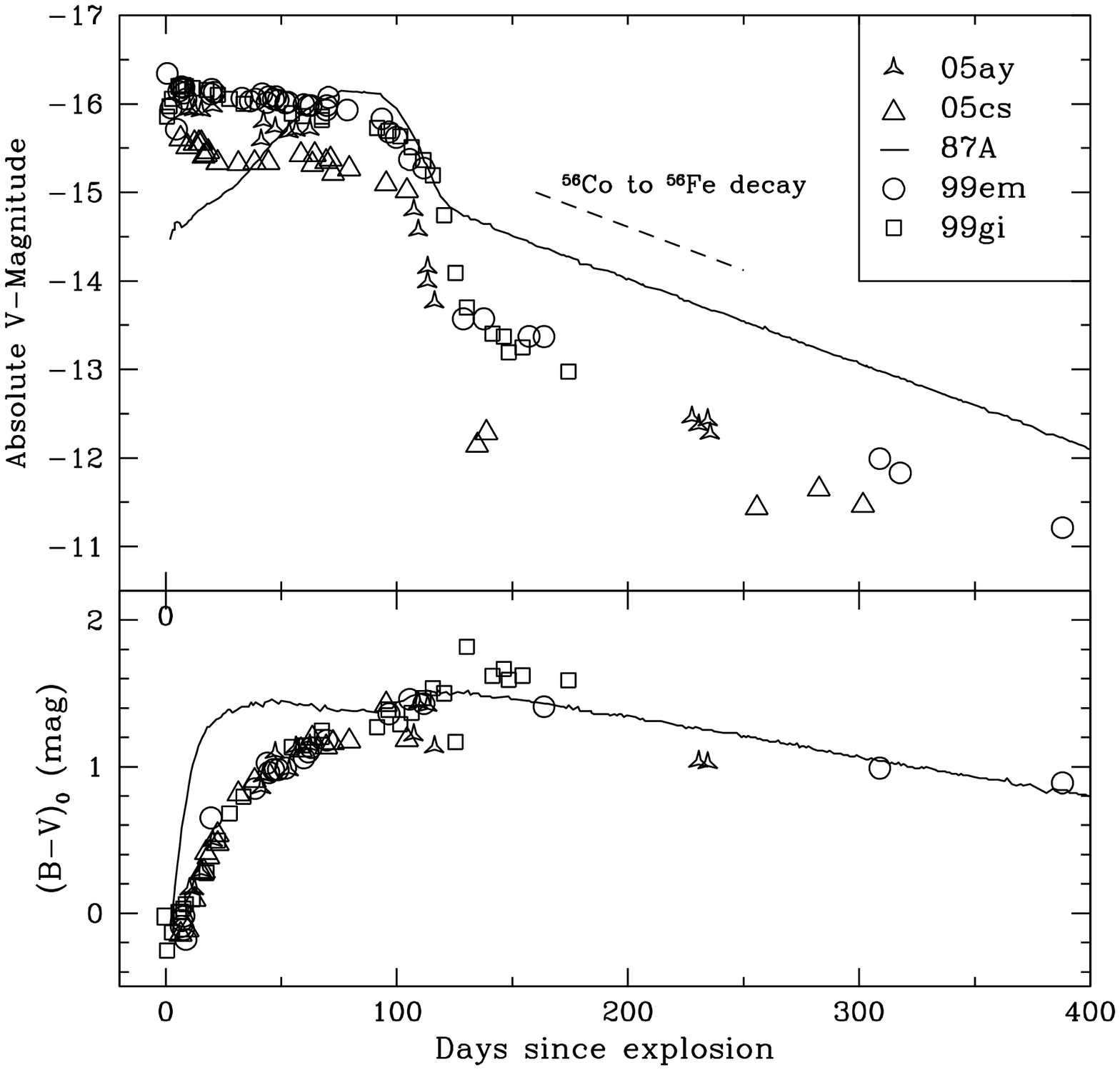}\includegraphics[height=.3\textheight]{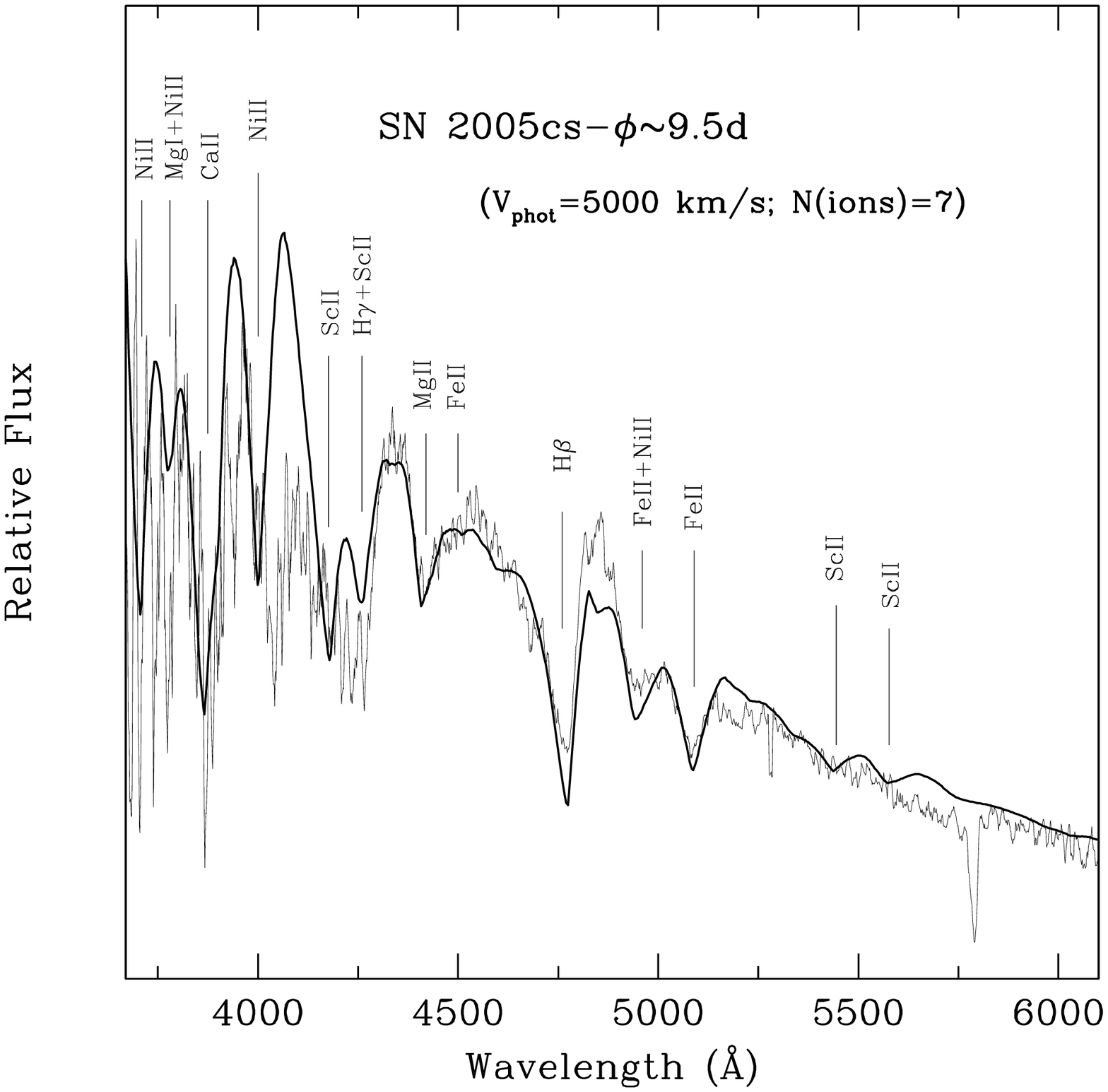}
  \caption{$Right$: The July 6 Spectrum of SN 2005cs, compared with the 
 SYNOW synthetic spectrum (thin line). Lines that are responsible for the most 
 conspicuous features are indicated.
 $Left$: Upper panel: the absolute V-light curves of SNe 2005ay (shorter 
 distance case) and 
 2005cs, compared to SNe 1987A ($D=50~ kpc, A_V^{tot}=0.6$ mag), 
 1999em ($D=8.8 ~Mpc; A_V^{tot}=0.31$ mag) 
 and 1999gi ($D=10.91 ~Mpc; A_V^{tot}=0.65$ mag). For SN 2005ay we plot the
 shorter distance case. Lower
 panel: the ``$(B-V)$'' intrinsic colour evolution of the SNe sample.}
\end{figure}
%%%%%%%%%%%%%%%%%%%%%%%%%%%%%%%%%%
 The absolute V-light curves of SNe 2005ay and 2005cs, 
 together with those of SNe 1987A, 
 1999em and 1999gi, are highlighted in Fig. 2 (top panel), while
 the ($B-V$) colour evolution is shown in the lower panel.
 In the following we summarize the main results of our analysis:\\
 A more detailed study of the two events has been presented recently 
 by Tsvetkov et al.(2006)\cite{Tsv06}.\\ \\
 {\bf The ejected $^{56}$Ni mass:}

 The radioactive decay phase of SN 2005cs is clearly not well 
 sampled, nevertheless
 the last two photometric points show a decline rate close 
 to the  $^{56}$Co decay. If we rely on these late data, adopting a 
 $^{56}$Ni mass of
 0.075 $M_{\odot}$ for SN 1987A \cite{Bouch93,Catch88}
 , a direct shift of the light curve tail of SN 1987A to fit late
 two points indicates a synthesized amount of $^{56}$Ni of 
 $\sim$0.017 $M_{\odot}$. The first 3 data of the radioactive tail
 may indicate an even lower $^{56}$Ni mass 
 ($\sim$9 $\times 10^{-3}M_{\odot}$)\\ 
 The observed data of SN 2005ay, around day 230 after 
 explosion, span about 8 days of observations. These late
 data are used as above to recover the 
 $^{56}$Ni mass (i.e. best fit with SN 1987A radioactive tail). This 
 method leads to $\sim$0.023 $M_{\odot}$ for the shorter distance
 and $\sim$0.051 $M_{\odot}$ for the larger distance.

 As a further check of the derived $^{56}$Ni masses, we make
 use of the well established correlation between the $^{56}$Ni mass 
 and the absolute $M_V$ at the plateau phase \cite{Elm03b,Ham03}.
 Using our estimates of $M_V$ during the plateau phase
 , i.e. $M_V$ $\simeq$ -15.33 for SN 2005cs and $M_V$ $\simeq$ -15.72; -16.61 
 for SN 2005ay , the method gives $^{56}$Ni masses of $\sim$0.018 
 $M_{\odot}$ for SN 2005cs and $\sim$0.026 $M_{\odot}$ for the shorter distance
  and $\sim$0.065 $M_{\odot}$ for the larger distance for SN 2005ay.
 This indicates that at least SN 2005cs
 belongs to the faint tail of the luminosity function of 
 type II SNe while the situation for SN 2005ay depends on the adopted
 distance.\\ \\
  {\bf The progenitor characteristics:}
%%%%%%%%%%%%%%%%%%%%%%%%%%%%%%%%%%
\begin{figure}
\includegraphics[height=6cm,width=13cm]{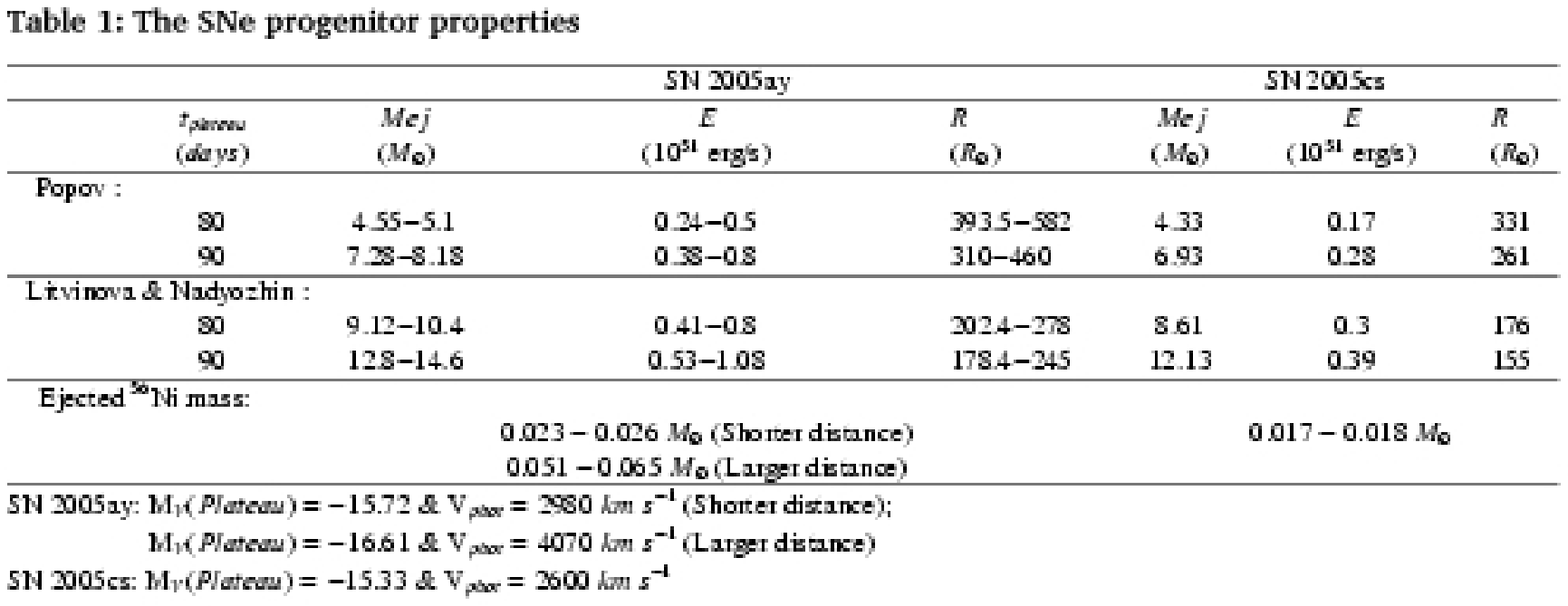}
%  \caption{}
\end{figure}
%%%%%%%%%%%%%%%%%%%%%%%%%%%%%%%%%%

 The type IIP-analytical model of Popov (1993)\cite{Pop93}, and the hydrodynamic models of 
 Litvinova $\&$ Nadezhin (1985)\cite{Lit85} are used to derive physical parameters 
 ($E$; $Mej$ and $R$). 
 The duration of the plateau ($tp$), the absolute V-magnitude 
 of the plateau ($M_V^p$) and the photospheric velocity ($Vp$) are the input 
 parameters.\\
 We evaluate the photospheric velocity at 50 days using the correlation:\\
  $V_{phot}(50d) \propto L_{plateau}^{0.464 \pm 0.017}$ \cite{Ham03}.
  We use these estimates, and we adopt a range of $tp=80-90$ days based on
 contraints by the early rise seen in both SN light curves. More precise 
 constraints on $tp$ should
 come from integrated bolometric light curves. The obvious declines seen 
 in $U$ and $B$ bands would affect the resulting bolometric light curves, 
 indicating shorter plateau than 
 seen in V light curves. The results are summarized in Table 1.
 We found the estimated mass range for SN 2005cs in agreement
 with limits established by using pre-supernova imaging \cite{Mau05}.

\end{document}